\documentclass[fleqn,useAMS,usenatbib]{mn2e}
\usepackage{epsfig,subfigure,amsmath}
\usepackage{color}

\definecolor{purple}{rgb}{1,0,1}

\newcommand{\lcdm}{$\Lambda$CDM}
\newcommand{\hmpc}{$h^{-1}$Mpc}

\newcommand{\beq}{\begin{equation}}
\newcommand{\eeq}{\end{equation}}
\def\ap{Alcock-Paczy\'{n}ski}

\bibpunct[; ]{(}{)}{;}{a}{}{,}

\title{A measurement of the Alcock-Paczy\'{n}ski effect using cosmic voids in the SDSS}

\author[P.~M.~Sutter, A.~Pisani, B.~D.~Wandelt, D.~H.~Weinberg]
{
\parbox{\textwidth}{
{P.~M. Sutter}$^{1,2,3}$ \thanks{Email: sutter@iap.fr},
Alice Pisani$^{1,2}$,
Benjamin D. Wandelt$^{1,2,4,5}$, and 
David H. Weinberg$^{6,3}$\\
}
\vspace{0.4cm}\\
\parbox[c]{\textwidth}{
$^{1}$ Sorbonne Universit\'{e}s, UPMC Univ Paris 06, UMR7095, Institut d'Astrophysique de Paris, F-75014, Paris, France \\
$^{2}$ CNRS, UMR7095, Institut d'Astrophysique de Paris, F-75014, Paris, France \\
$^{3}$ Center for Cosmology and AstroParticle Physics, Ohio State University, Columbus, OH 43210\\
$^{4}$ Department of Physics, University of Illinois at Urbana-Champaign, Urbana, IL 61801\\
$^{5}$ Department of Astronomy, University of Illinois at Urbana-Champaign, Urbana, IL 61801\\
$^{6}$ Department of Astronomy, Ohio State University, Columbus, OH 43210
}}

\begin{document}

\maketitle

\label{firstpage}

\begin{abstract}
We perform an \ap~test using stacked cosmic voids identified
in the SDSS Data Release 7 main sample and Data Release 10
LOWZ and CMASS samples. We find $\sim$1,500 voids out to redshift $0.6$ 
using a heavily modified and extended version of the watershed algorithm 
{\tt ZOBOV}, which we call {\tt VIDE} (Void IDentification and Examination).
To assess the impact of peculiar velocities we use the mock void catalogs 
presented in Sutter et al. (2013). 
We find a constant uniform flattening of 14\% along the line of sight 
when peculiar velocities are included. 
This flattening appears universal for all void sizes at all redshifts 
and for all tracer densities.
We also use these mocks to identify an optimal stacking strategy. 
After correcting for systematic effects
we find that our \ap~measurement leads to a preference of our best-fit 
value of $\Omega_{\rm M}\sim 0.15$ over $\Omega_{\rm M} = 1.0$ by a 
likelihood ratio of 10. 
Likewise, we find a factor of $4.5$ preference of the likelihood ratio for 
a \lcdm~$\Omega_{\rm M} = 0.3$ model and a null measurement.
Taken together, we find substantial evidence for the \ap~signal in our
sample of cosmic voids.
Our assessment using realistic 
mocks suggests that measurements with future SDSS releases and other surveys 
will provide tighter cosmological parameter constraints.
The void-finding algorithm and catalogs used in this work will be made 
publicly available at http://www.cosmicvoids.net.
\end{abstract}


% -----------------------------------------------------------------------------
% -----------------------------------------------------------------------------
\section{Introduction}

Alternative cosmological probes offer complementary and orthogonal 
avenues for answering pressing questions such as the nature of dark energy 
and the growth of large-scale structure (for a recent review of traditional 
and alternative probes, see~\citealt{Weinberg2013}).
One such alternative probe is the \ap~(AP) test~\citep{Alcock1979}, which 
instead of using standard candles such as 
Type Ia supernovae~\citep[e.g.,][]{AlderingGreg2002} 
or standard rulers such as baryon acoustic 
oscillations (BAO;~\citealt{Anderson2012}) 
relies on standard \emph{spheres} for a geometrical test of cosmological 
parameters. 
The principal concept underlying the AP test is simple: 
in a properly chosen cosmology, spheres will maintain a uniform 
ratio of line-of-sight to angular extent. Deviations from sphericity 
as a function of redshift thus reveal the true cosmology. 

Since the AP test relies only on statistical isotropy, it has been 
considered for --- and applied to --- a variety of systems and 
features, such as 
the Lyman-$\alpha$ forest~\citep{Hui1999, McDonald1999,EriksenK.A.2005}, 
the power spectrum of the epoch of reionization~\citep{Nusser2005},
galaxy cluster autocorrelation spectra~\citep{Kim2007a},
galaxy clustering in the WiggleZ survey~\citep{Blake2011},
galaxy pairs~\citep{Jennings2012},
and the BAO feature in the Sloan 
Digital Sky Survey~\citep{Reid2012}.

Current successful applications of the AP 
test~\citep[e.g.,][]{Blake2011, Reid2012} are limited to large 
scales, fundamentally limiting their constraining power. 
A promising way to apply the AP test to smaller 
scales (thereby reducing statistical uncertainties) 
while still avoiding large systematics
is to use cosmic voids~\citep{LavauxGuilhem2011}, 
the large underdense regions in the cosmic 
web~\citep{Thompson2011}. 
Voids offer potentially revolutionary potential:~in terms of 
statistical power \citet{LavauxGuilhem2011} 
predicted that the AP test applied with voids in 
the upcoming Euclid survey~\citep{Laureijs2011} will outperform 
BAO in constraining dark energy equation of state parameters 
 by up to a factor of ten.

The power of voids comes from two aspects. First, their small size 
compared to the BAO feature, down to $\sim$5~\hmpc~\citep{Sutter2012a},
gives them significant statistical weight. Second, since their 
evolution is in the linear or quasi-linear regime, systematic effects 
due to peculiar velocities will be highly suppressed and easy 
to model out. Even though the BAO feature is more linear than 
void features, voids are a collective phenomenon defined by many galaxies 
and so void profiles (and shapes) are a 
cross-correlation~\citep{Hamaus2013}, whereas the BAO relies on 
galaxy-galaxy auto-correlation. 
As we will see in this paper, the cross-correlation has potential to be 
less affected by peculiar velocities and other systematics than the 
auto-correlation. 

Put simply, voids are simple objects. 
For example,~\citet{Hamaus2014} discovered a single two-parameter 
density profile that describes voids of all sizes, 
and~\citet{Sutter2013a} applied this profile to reveal that voids 
in theory (e.g., in dark matter simulations) obey self-similar 
scaling relations to voids in observations (e.g., in galaxy surveys).
Also, the merger tree analysis of~\citet{Sutter2014} found that voids 
essentially do not move and do not merge over their lifetimes; 
evolutionary dynamics do not overwhelm primordial cosmological 
information.

Even though current galaxy redshift surveys are not optimized 
for finding large numbers of voids (due to their relative sparsity, 
low redshift, and complicated survey geometries), cosmological 
measurements with voids are still possible: for example, the largest 
publicly-available void catalog\footnote{http://www.cosmicvoids.net}
~\citep{Sutter2012a, Sutter2013c} has enabled observations such 
as the 
ISW effect~\citep{Planck2013b} and gravitational 
anti-lensing~\citep{Melchior2013}. 
Previously, we applied the AP test methodology described 
in~\citet{LavauxGuilhem2011} to voids found in the SDSS 
DR7 main sample and LRG catalogs~\citep{Sutter2012a}, 
but due to the small number of voids 
found no statistically significant result~\citep{Sutter2012b}.

In this work we extend the void AP analysis to higher redshifts and 
to more voids using the BOSS Data Release 10 LOWZ and CMASS galaxy 
catalogs~\citep{Ahn2014}. We also use mock catalogs tuned to our 
observational surveys to examine 
the systematic impact of peculiar velocities noted 
in the pure $N$-body simulations of~\citet{LavauxGuilhem2011}. 
We use these mocks to find an optimal binning strategy to increase sensitivity to
our potential signal and then use the resulting size bins on the data.

In Section~\ref{sec:samples} we describe the galaxy samples and 
void catalogs to be used in the AP analysis. We discuss our 
method for measuring distortions in void shapes and the application 
to an AP test in Section~\ref{sec:method}. Section~\ref{sec:systematics} 
focuses on systematics due to peculiar velocities, while 
Section~\ref{sec:optimization} features an analysis of our strategy 
for optimizing the signal given the limited number of voids in our current void catalog. We present our AP results in Section~\ref{sec:results} 
and offer prospectives on future work in Section~\ref{sec:conclusions}.

% -----------------------------------------------------------------------------
% -----------------------------------------------------------------------------
\section{Galaxy \& Void Samples}
\label{sec:samples}

For each galaxy we transform its sky latitude
$\theta$, sky longitude $\phi$, and 
redshift $z$ into a comoving coordinate system:
\begin{eqnarray*}
  x' & = & D_c(z) \cos{\phi} \cos{\theta}, \\
  y' & = & D_c(z) \sin{\phi} \cos{\theta}, \\
  z' & = & D_c(z) \sin{\theta}, 
\label{eq:transform}
\end{eqnarray*}
where $D_c(z)$ is the comoving distance to the galaxy
at redshift $z$.
We assume cosmological parameters consistent with a \lcdm~cosmology 
as given by the WMAP 7-year
results~\citep{Komatsu2011}: $\Omega_{\rm M}=0.3$, $\Omega_\Lambda=0.7$,
and $h=0.71$.
We make no corrections for peculiar velocities.

We construct volume-limited galaxy populations by making cuts in magnitude 
and redshift from the SDSS Data Release 7 main sample~\citep{Abazajian2009} 
and SDSS-III BOSS Data Release 10~\citep{Ahn2014} LOWZ and CMASS 
galaxy catalogs. We make these cuts after applying evolution and 
$K$-corrections and computing absolute magnitudes using the above
WMAP 7-year cosmological parameters. The main sample 
cuts are described more fully in~\citet{Sutter2012a}, and we split the 
LOWZ catalog into four samples starting at $z=0.1$ and the CMASS 
catalog into two samples starting at $z=0.5$. To avoid overlap with the 
DR7 samples we ignore the first lowest-redshift LOWZ sample. 

Our void-finding procedure as applied to observational 
data is described in detail 
in~\citet{Sutter2012a} and~\citet{Sutter2013c}.
Briefly, we use a heavily modified version 
of {\tt ZOBOV}~\citep{Neyrinck2008, LavauxGuilhem2011, Sutter2012a, 
Sutter2013a, Sutter2014}, which we call {\tt VIDE}, for Void IDentification 
and Examination. This approach uses a Voronoi tessellation 
to construct a density field from the galaxy population.
Our  construction of volume-limited samples ensures that 
we have uniform density across the void surface so that we do not need 
to include any weighting in the tessellation step.
This density field has topological features such as basins and 
ridgelines, and {\tt VIDE} assembles adjacent basins into voids using 
a watershed algorithm. Thus a void is simply an arbitrarily-shaped 
depression in the density field bounded by high-density ridgelines.
{\tt VIDE} does not apply any additional smoothing before 
the watershed step. In addition to the above references, 
we will discuss {\tt VIDE} in detail in a forthcoming paper.

A void can have any mean and minimum density, since the watershed 
algorithm includes as member particles galaxies in the surrounding high-density 
walls.
However, we place a restriction such that the \emph{walls} between 
adjacent basins cannot be merged into a larger void unless 
the density of that wall is less than $0.2 \bar{\rho}$. This 
prevents the growth of voids deeply into clusters~\citep{Neyrinck2008}.
Also, we remove voids with central densities greater than 
$0.2 \bar{\rho}$, measured within $1/4$ of the effective void 
radius. This cleaning aids in the shape measurement below. We only 
include voids with effective radii greater than the mean galaxy 
separation. 

Throughout this work we will refer to the void radius. We define 
the effective radius to be the radius of a sphere with the 
same volume as the void, where the void volume is the sum of all 
the Voronoi cell volumes that comprise the void.

To accommodate the survey boundaries and masks we place a large 
number of mock particles 
along any identified edge. These mock particles have essentially 
infinite density and thus prevent the watershed from growing voids 
outside the survey area. 
After finding voids we take the \emph{central} selection, 
where voids are guaranteed to sit well 
away from any survey boundaries or internal holes: the maximum 
distance from the void center to any member particle is less than the 
distance to the nearest boundary. As discussed in~\citet{Sutter2012a} 
and~\citet{Sutter2012b}, this ensures a fair distribution of 
void shapes and alignments.

The voids identified with this approach in the DR7 main sample 
 are already publicly available.

Table~\ref{tab:samples} summarizes the names of our volume-limited samples, 
the maximum absolute magnitude, redshift bounds, mean galaxy number 
density, and the total number of voids identified in those samples.

\begin{table}
\centering
\caption{Volume-limited galaxy samples used in this work.}
\tabcolsep=0.11cm
\footnotesize
\begin{tabular}{cccccc}
  Sample Name & $M_{r, {\rm max}}$ & $z_{\rm min}$ &
              $z_{\rm max}$ & $\bar{n}^{-1/3}$ & 
              $N_{\rm void}$ \\
   &  &  & & (\hmpc) &\\
  \hline  \hline
dr72dim1 & -18.9 & 0.0 & 0.05 & 3.5 & 52 \\
dr72dim2 & -20.4 & 0.05 & 0.1 & 4.9 & 174 \\
dr72bright1 & -21.4 & 0.1 & 0.15 & 7.4 & 183 \\
dr72bright2 & -22.0 & 0.15 & 0.2 & 12.5 & 96 \\
\\
dr10lowz2 & -19.5 & 0.2 & 0.3 & 13.8 & 137 \\
dr10lowz3 & -20.0 & 0.3 & 0.4 & 15.1 & 199 \\
dr10lowz4 & -20.5 & 0.4 & 0.45 & 16.4 & 91 \\
\\
dr10cmass1 & -19.5 & 0.45 & 0.5 & 14.0 & 230 \\
dr10cmass2 & -19.5 & 0.5 & 0.6 & 15.2 & 697 \\
\hline
\end{tabular}
\label{tab:samples}
\end{table}

% -----------------------------------------------------------------------------
% -----------------------------------------------------------------------------
\section{Method}
\label{sec:method}

The \ap (AP) test uses a set of 
standard spheres to measure cosmological 
parameters by taking the ratio of line-of-sight distances to 
angular diameters:
\begin{equation}
  \frac{\delta z}{\delta d} = \left( \frac{H_0}{c} \right)^2 
                               \frac{D_A(z) E(z)}{z},
\label{eq:ap}
\end{equation}
where $\delta z$ is an extent along the line of sight, 
$\delta d$ is an angular extent, $H_0$ is the Hubble constant, 
$D_A(z)$ is the angular diameter distance at redshift $z$, and 
$E(z)$ is the expansion rate at that redshift. For this work we will 
assume a flat \lcdm~universe, and thus $D_A(z)$ becomes
\begin{equation}
  D_A(z) = \frac{c}{H_0} \int_0^z \frac{d z'}{E(z')},
\label{eq:angdiam}
\end{equation}
with
\begin{equation}
  E(z) = \left( 
                 \Omega_m (1+z)^3 + \Omega_\Lambda
                 \right)^{1/2}.
\label{eq:hubble}
\end{equation}
In the above, $\Omega_m$ and $\Omega_\Lambda$ are, respectively, 
the present-day matter and dark
energy densities relative to the critical density.

We define the \emph{stretch} parameter $e_v(z)$ as
\begin{equation}
  e_v(z) \equiv \frac{c}{H_0} \frac{\delta z}{\delta d}.
\label{eq:stretch}
\end{equation} 

While a single void is hardly a standard sphere and inappropriate for the 
AP test~\citep{Ryden1995}, in an isotropic universe voids have no preferred 
orientation: a \emph{stack} of voids will be a 
sphere~\citep{LavauxGuilhem2011}. 
To stack voids we align their barycenters, which are the volume-weighted
centers of all the Voronoi cells in each void:
\begin{equation}
  {\bf X}_v = \frac{1}{\sum_i V_i} \sum_i {\bf x}_i V_i,
\label{eq:barycenter}
\end{equation}
where ${\bf x}_i$ and $V_i$ are the positions and Voronoi volumes of
each tracer $i$, respectively.
As we stack, we align the voids so that they share a common line of sight.
This same approach has been used to construct 
real-space density profiles using projections~\citep{Pisani2013}.
We do not apply any rescalings to the void sizes as we stack.

Thus we can measure the stretch $e_v(z)$ 
of stacked voids within independent redshift slices. 
We placed our galaxy samples into a comoving coordinate system 
assuming a \lcdm~cosmology. If this is the correct cosmology then 
the AP test will return unity for all redshifts: $e_v(z)=1$. Deviations 
from unity as a function of redshift will depend on cosmological 
parameters (Eq.~\ref{eq:ap}). 

To measure the stretch of a stacked void we take all galaxies within 
a radius cutoff of $2 R_{\rm max}$, where $R_{\rm max}$ is the maximum 
void size in the stack. We project these galaxies onto a two-dimensional 
plane,
\begin{eqnarray}
  d_v & = & \sqrt{x_{\rm rel}^2 + y_{\rm rel}^2} \\
  z_v & = & |z_{\rm rel}| \nonumber,
\label{eq:transformation}
\end{eqnarray}
where $(x_{\rm rel}, y_{\rm rel}, z_{\rm rel})$ are the galaxy coordinates
relative to the void barycenter ${\bf X}_v$:
\begin{equation}
  {\bf x}_{\rm rel} \equiv {\bf x}' - {\bf X}_v.
\end{equation}
We then transform the line-of-sight coordinate $z_v$ by a factor 
$e_v$ ($z_v' \equiv e_v z_v$) until the ellipticity measured within a 
sphere of radius $0.7 R_{\rm max}$ is unity. In testing we found 
this radius to provide the best balance between gathering as many 
galaxies as possible for high signal-to-noise while avoiding fluctuations
outside the void proper. We will validate this approach 
in the section below.
We calculate the ellipticity $\epsilon$ via the inertia tensor:
\begin{equation}
  \epsilon = \sqrt{ \frac{2 \sum z_{v,i} }{ \sum d_{v,i} } },
\label{eq:ellipticity}
\end{equation}
where the sums are taken over all galaxies within $0.7 R_{\rm max}$.

\begin{figure}
  \centering
  {\includegraphics[type=png,ext=.png,read=.png,width=\columnwidth]{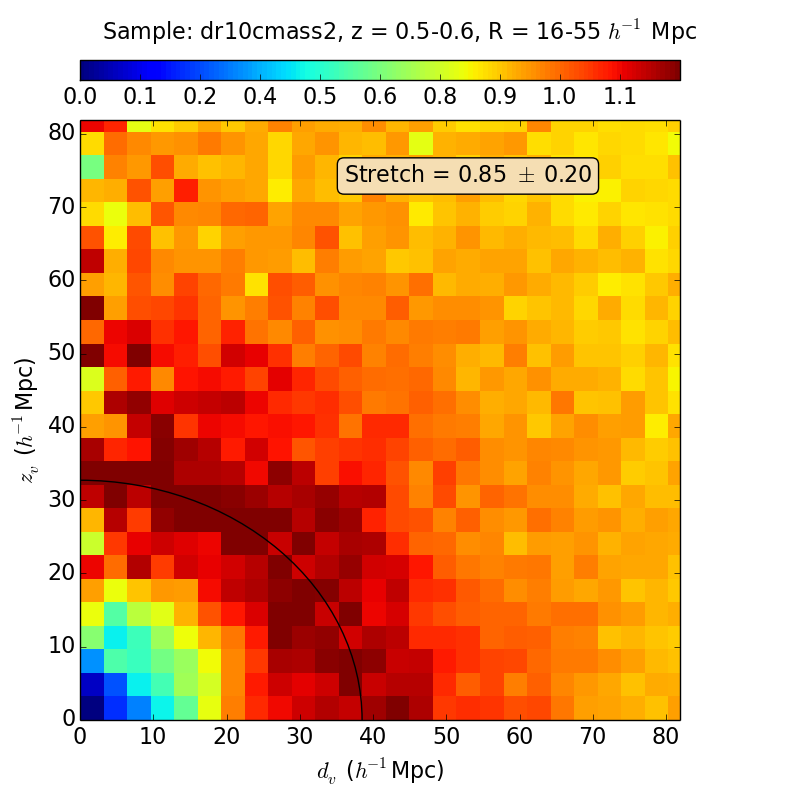}}
  \caption{
           Example of a void stack in two dimensions 
           and its stretch measurement. This stack includes 
           voids from size 16 to 55~\hmpc. For clarity 
           void galaxies are binned onto a uniform grid with cell 
           size $\sim 3$~\hmpc, though the bins are not used in the 
           stretch calculation. The black line shows an ellipse with 
           the same stretch as the stacked void,
           and the caption shows the measured stretch and 
           uncertainty. The uncertainty is the $1\sigma$ deviation taken
           from 1,000 bootstrapped samples.  
           The color bar shows the density in each grid cell in units of the 
           mean density of this sample.
          }
\label{fig:stack}
\end{figure}

We identify the necessary rescaling factor $e_v$ as the void stretch 
which enters into the AP measurement above.
To estimate our uncertainty for each stack we repeat the above process for 
1,000 bootstrap samples.
Figure~\ref{fig:stack} shows an example void stack taken from the CMASS 
sample and the measured stretch using this approach.

% -----------------------------------------------------------------------------
% -----------------------------------------------------------------------------
\section{Analysis of Systematics}
\label{sec:systematics}

The theoretical 
analysis of the AP effect with voids discussed in~\citet{LavauxGuilhem2011} 
revealed the presence of systematic 
effects due to peculiar velocities: a uniform compression along the 
line of sight for voids at all redshifts. In order to recover the 
expected AP signal, a single correction factor 
of $1.16$ (with an observed flattening of 14\%, the correction factor 
is $1/(1 - 0.14) = 1.16$) had to be applied.
However, that analysis focused on a relatively thin stack of voids 
($8-9$~\hmpc) in a dark matter simulation. We extend this work 
to examine the impacts of peculiar velocities on more realistic 
galaxy populations.

For our study we take the publicly-available void population 
presented in~\citet{Sutter2013a}; namely, the \emph{Halos Dense} and 
\emph{Halos Sparse} samples. These two void samples are drawn from 
$N$-body simulation 
halo catalogs with similar number densities (and, by construction, clustering 
properties) of a high-density galaxy survey such as the DR7 main sample 
and a low-density survey such as DR10 CMASS, respectively. 
Note that we do not take the void catalogs drawn
 from an Halo Occupation Distribution (HOD;~\citealt{Berlind2002})  
galaxy population due to 
the ambiguity in applying HOD modeling at higher redshifts.
However, as~\citet{Sutter2013a} noted, the void populations between 
halos and galaxies are almost 
indistinguishable for these kinds of analysis, and thus the halos provide a good 
proxy for the galaxy populations.
The particle positions were perturbed according to their peculiar velocities 
before finding voids.

We analyze three simulation snapshots at redshifts $z=0.05$, $0.25$, 
and $0.67$, 
and subdivide each snapshot into four slices in the $z$-direction.
These slices each have a comoving width of $300$~\hmpc, or $\Delta z \sim 0.1$. 
In total there are $\sim 25,000$ voids in the \emph{Halos Dense} 
catalogs and $\sim 5,000$ in the \emph{Halos Sparse} catalogs.
For the \emph{Halos Dense} catalogs, we construct stacks of width 
$\Delta R = 10$~\hmpc. The fewer number of voids in the 
\emph{Halos Sparse} catalogs necessitates the use of wider stacks with 
$\Delta R = 40$~\hmpc. We start adding voids to stacks beginning 
with voids with effective radii equal to the mean particle separation.
With these stacks we applied the above AP analysis assuming the 
$z$-direction to be the line of sight. 

\begin{figure*}
  \centering
  {\includegraphics[type=png,ext=.png,read=.png,width=0.48\textwidth]{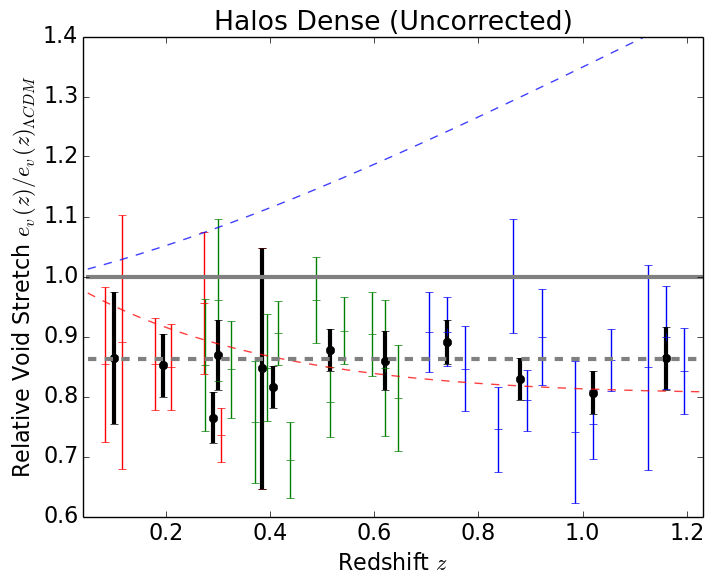}}
  {\includegraphics[type=png,ext=.png,read=.png,width=0.48\textwidth]{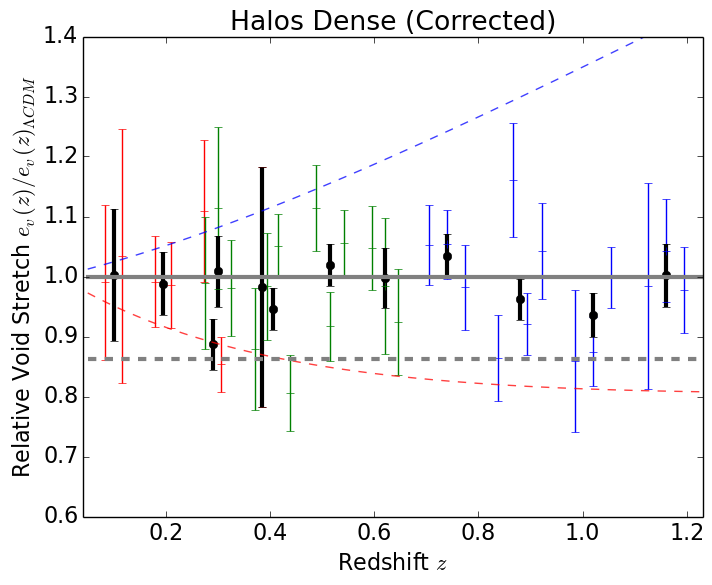}}
  {\includegraphics[type=png,ext=.png,read=.png,width=0.48\textwidth]{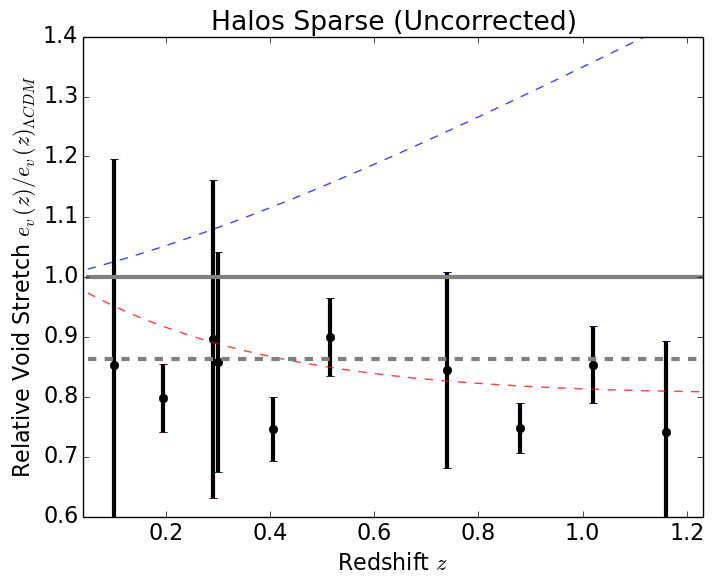}}
  {\includegraphics[type=png,ext=.png,read=.png,width=0.48\textwidth]{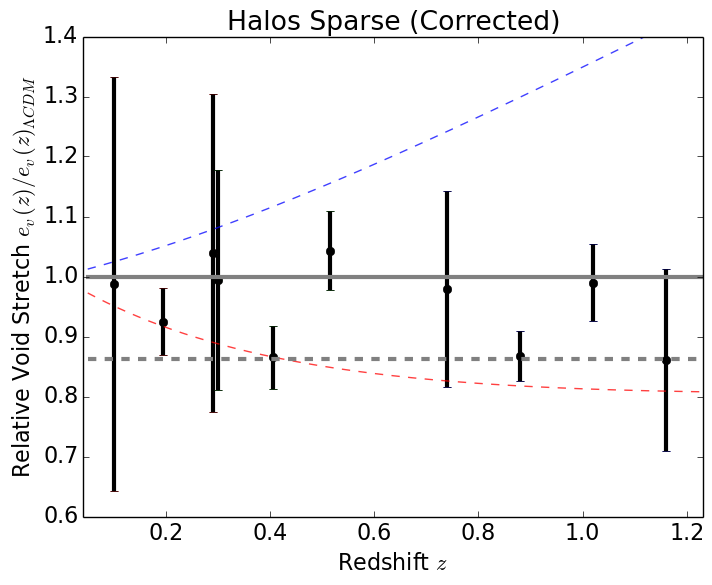}}
  \caption{
           Stretch as a function of redshift ($e_v(z)$) relative to the stretch
           expected in a~\lcdm~cosmology for the mock void samples. Since 
           the mocks are drawn from a~\lcdm~universe, we expect a stretch
           measurement of unity, as indicated by the solid grey horizontal  
           line. Peculiar velocities induce a uniform flattening, and the 
           dashed grey line shows the offset found in the dark matter 
           analysis of~\citet{LavauxGuilhem2011}. Each colored point 
           with error bar is the stretch measurement for a single stack 
           (e.g., $10-15$~\hmpc), 
           while the thick black points with error bars are the weighted mean 
           measurements in a redshift slice (if there is only a single stack
           in the slice we only show the black point with error bar). 
           The individual stack measurements are colored by redshift slice.
           Different color points correspond 
           to particular simulation snapshots. 
           For reference the dashed blue (red) line corresponds to the 
           relative stretch for an $\Omega_M = 0.0$ ($1.0$) universe. 
           The top panels are from the \emph{Halos Dense} mocks 
           and the bottom panels are from the \emph{Halos Sparse} mocks.
           The left-hand panels are the raw measurement; the right-hand panels 
           show the measurement after applying a uniform correction factor
           of $1.16$.
          }
\label{fig:mocks_hubble}
\end{figure*}

Figure~\ref{fig:mocks_hubble} shows the measured stretch as a function 
of redshift, $e_v(z)$, for these mock setups. Since the voids are 
drawn from an $N$-body simulation with~\lcdm~parameters, and 
we are assuming the same parameters for the AP test, we expect 
no stretch. Instead, we see the same uniform 14\% line-of-sight 
flattening, and hence a necessary $e_v(z)$ correction factor of $1.16$, as 
in~\citet{LavauxGuilhem2011}. We indicate this in the figure 
with a horizontal 
grey dashed line. We find, but do not show, the expected value 
of $e_v(z) = 1.0$ when repeating this analysis on voids without 
including peculiar velocities.

We see that the systematic flattening due to peculiar velocities first observed 
by~\citet{LavauxGuilhem2011} persists for voids in mock galaxy populations, 
and also persists for all void sizes at all redshifts. There is some indication
 of a slight deviation for the highest redshifts in the \emph{Halos Sparse} 
voids, but this is outside the redshift range that we consider in data.
These results are not necessarily surprising: the recent works 
of~\citet{Hamaus2014} and~\citet{Sutter2013a} have shown that a single 
universal density profile applies to voids in all tracer populations 
and enables the definition of a scaling relationship among them. 
Thus it is plausible that systematic effects that 
impact voids in simulation will 
be similar in real galaxy populations. 

We leave a detailed study of the cause of this systematic offset to 
future work: a better understanding of this effect can be reached by directly 
comparing on a one-to-one basis voids in mocks with and without 
peculiar velocities included (similar to the approach taken 
in~\citealt{Sutter2013b} to examine the impacts of galaxy sparsity and bias).
A preliminary study shows that peculiar velocities impart a 
uniform flattening to void shapes along the line of sight. 
While one might expect peculiar velocities to elongate 
individual voids~\citep{Ryden1996}, the situation is less clear in 
a statistical sample of the full cosmic web: the elongation in 
the dynamically outflowing portion of voids competes 
with a thickening of the walls separating them. 
In addition, peculiar velocities will have an effect on the 
assignment of a given portion of survey volume to individual voids. 

% -----------------------------------------------------------------------------
% -----------------------------------------------------------------------------
\section{Optimization}
\label{sec:optimization}

We may use the measured flattening as a metric for optimization: we wish 
to have as many independent stacks as possible, but too few voids --- 
and hence tracers --- in a 
stack will degrade the shape measurement, leading to $e_v=1.0$ even in the 
presence of peculiar velocities. 
This occurs because Poisson fluctuations overwhelm the shape distortions 
caused by~\ap~stretching and peculiar velocities.
We investigated the minimum number 
of voids required, the minimum and maximum 
stack width, the minimum void size to start stacking, 
the radius at which the ellipticity 
is calculated, whether to use all particles in the stack or 
only particles that are members of a {\tt VIDE} void, and whether 
to rescale voids to the same radius or not.

First, we found that rescaling voids severely degrades the measurement 
due to the broadening of the high-density wall (for example, see the density 
profiles in~\citealt{Sutter2012b}). This broadening occurs because galaxies 
are moved to larger radii, leaving fewer galaxies within the void 
to use for the shape measurement and thereby increasing the Poisson noise.  
Also, we are able to reliably measure 
shapes when restricting ourselves to void-galaxies only, but the minimum 
number of voids in a stack is necessarily larger, since in this case we 
are making measurements with fewer numbers of particles per void.
We found the radius choice of $0.7 R_{\rm max}$ to be the most robust: 
at smaller radii we lose too many galaxies, and larger radii include 
fluctuations outside the void wall in the shape estimation, leading 
to highly variable measurements from stack to stack.
When making broad stacks, as we will see below, we found that this criterion 
still succeeds even when some voids are smaller 
than $0.7 R_{\rm max}$ of the stack. Figure~\ref{fig:stack} shows why 
this works: in a stacked void without rescaling, small voids contribute 
to the inner portions of the wall, while larger voids add galaxies 
to the outer edges. Combined they give a very broad, smooth feature 
that robustly measures the overall shape.

We found for both catalogs that we may begin stacking 
at the minimum void size provided by {\tt VIDE}, 
which is the mean particle separation. 
We are able to recover reliable stretch measurements when 
setting larger thresholds for the minimum void size, 
but with larger statistical uncertainties. Including all voids 
down to the mean particle separation did not introduce any 
systematic error.

For the \emph{Halos Dense} voids we require at least $\sim$50 voids 
per stack to preserve the shape information, while for the \emph{Halos Sparse}
voids we must use at least $\sim$150 voids per stack. Relatedly, 
the minimum stack width for reliable measurements in 
\emph{Halos Dense} was $\sim$5~\hmpc, while in 
\emph{Halos Sparse} was $\sim$20~\hmpc. We found maximum 
reliable stack widths of $\sim$20~\hmpc\ and $\sim$40~\hmpc\ for the 
\emph{Halos Dense} and \emph{Halos Sparse} mocks, respectively. 
The measurement degrades for very wide stacks because we increase 
the radius at which we compute ellipticities while adding 
relatively few new voids to the stack.

In addition to the method of general measurement of 
ellipticity in the stack, we repeated the above analysis 
for the MCMC shape-fitting algorithm 
presented in~\citet{LavauxGuilhem2011} and applied to data 
in~\citet{Sutter2012b}, using both a cubic density profile and the 
universal density profile of~\citet{Hamaus2014}. With both profiles 
we are able to reproduce these AP measurements, but we require 
a factor of $\sim$2 greater number of voids per stack 
to avoid catastrophic failures of the estimator. 
Also, voids rescaling is required for the profile-fitting method to 
avoid being dominated by large fluctuations inside the void wall, 
and rescaling can only be reliably applied for relatively narrow 
radius bins, as seen above.

While the methods based on profile fitting achieve much tighter 
constraints once the number of voids is above this threshold, we 
find ellipticity-based methods such as the one described 
in Section~\ref{sec:method} more robust (i.e., less subject to 
Poisson noise) for the number of 
voids available in the extant sample. Also, with relatively 
few voids available in data we are forced to use wide radial bins, 
where the profile-fitting method is less robust.
We will comment in the Conclusions on the comparison of our results 
and the forecasts of the 
AP measurement based on profile-fitting shape measurements.

Finally, we also repeated this analysis in redshift space using 
the coordinate system of~\citet{LavauxGuilhem2011} and~\citet{Sutter2012b} 
and found identical results.

% -----------------------------------------------------------------------------
% -----------------------------------------------------------------------------
\section{Results}
\label{sec:results}

From the above optimization analysis, we selected the stacks 
listed in Table~\ref{tab:stacks}. This table lists the sample name, 
the void size range in the stack, and the number of voids in the stack. 
Within each sample we attempted to create as many stacks as possible 
given the minimum required number of voids in order to correctly 
estimate the expected stretch in the mocks. We used the \emph{Halos Dense} 
mock as a guide for the DR7 main sample, and the \emph{Halos Sparse} 
mock for guidance with the DR10 LOWZ and CMASS samples.  
We discard the \emph{dr72dim1} and \emph{dr10lowz4} samples due to a lack 
of a sufficient number of voids.
We have verified in mocks that our chosen bin sizes and typical 
number of voids per stack are able to capture the correct AP 
measurement (indeed, the stacks used in Figure~\ref{fig:mocks_hubble} 
are capped to 50 and 200 voids for the \emph{Halos Dense} and 
\emph{Halos Sparse} samples, respectively). 

\begin{table}
\centering
\caption{Void stacks used in the analysis.}
\tabcolsep=0.11cm
\footnotesize
\begin{tabular}{ccc}
  Sample Name & $R_{\rm min}$ - $R_{\rm max}$ (\hmpc) &
              $N_{\rm voids}$ \\
  \hline  \hline
dr72dim2 & 5 - 8 & 78 \\
          & 8 - 12 & 59 \\
dr72bright1 & 7 - 12 & 88 \\
            & 12 - 20 & 71 \\
dr72bright2 & 12 - 28 & 75 \\
\\
dr10lowz2 & 14 - 55 & 135 \\
dr10lowz3 & 15 - 55 & 195 \\
\\
dr10cmass1 & 14 - 55 & 219 \\
dr10cmass2 & 16 - 55 & 659 \\
\hline
\end{tabular}
\label{tab:stacks}
\end{table}

\begin{figure}
  \centering
  {\includegraphics[type=png,ext=.png,read=.png,width=\columnwidth]{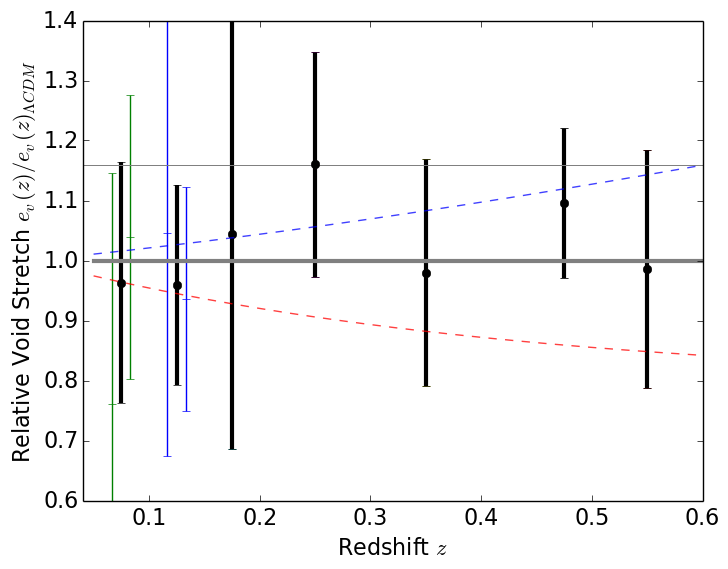}}
  \caption{
           Stretch as a function of redshift relative to the expected 
           stretch in a~\lcdm~cosmology (thick solid grey horizontal line) 
           and a null measurement (thin solid grey horizontal line).
           Each black point with error bars 
           corresponds to the weighted mean measurement in each galaxy 
           sample, while colored points with error bars show individual 
           stack measurements in a single sample, if available.
           The measurements are already corrected for peculiar velocities
           using a single $1.16$ correction factor, as discussed 
           in Section~\ref{sec:systematics}.           
           Error bars are $1\sigma$ uncertainties obtained using 1,000
           bootstrapped samples.
           For reference the dashed blue (red) line corresponds to the
           relative stretch for an $\Omega_M = 0.0$ ($1.0$) universe.
          }
\label{fig:drall_hubble}
\end{figure}

Figure~\ref{fig:drall_hubble} shows our void stretch measurement using the 
above stacks. In this figure we have already applied the uniform 
$1.16$ correction factor 
discussed above. We see that after this correction is applied 
our measurements scatter around the expected value of $e_v(z)=1.0$. These error bars 
are larger than the mean measurements shown in 
Figure~\ref{fig:mocks_hubble}, since in mocks we have many more 
individual stacks so that the mean measurement has relatively smaller 
uncertainty. With our limited number of voids in data we have, at 
best, two stacks per sample. 

Note that if our shape measurements were degraded due to an insufficient number 
of voids in each stack, we would have measured a signal consistent 
with $e_v(z)=1.0$ \emph{before} 
correcting for peculiar velocities, and $e_v(z)=1.16$ 
after corrections, since our measurement will be 
dominated by Poisson noise, as discussed above. 
For example, the \emph{dr10lowz2} sample has the fewest number 
of voids for the given sparsity and is the most discrepant from 
the expected measurement.
This serves as a useful null 
test. To evaluate the significance of our results compared 
to such a null measurement, we evaluate the likelihood ratio
\begin{equation}
  K \equiv \frac{ \mathcal{L}_{ {\rm \Lambda CDM}}}{\mathcal{L}_0} = \exp \left[ 
   -\frac{1}{2} \left( \chi_{ {\rm \Lambda CDM}}^2 - \chi_0^2 \right)
   \right],
\label{eq:ratio}
\end{equation}
where a subscript of \lcdm~refers to the expected measurement and a
 subscript of $0$ for a null measurement, and $\chi^2$ is the 
residual sum-of-squares. 
To calculate the likelihood ratio we compute the $\chi^2$ of 
our measurements against the expected \lcdm~model (a constant value of 
unity for all redshifts) and again against the expected 
null result (a constant value of $1.16$ for all redshifts). The 
likelihood ratio measures the amount by which our data prefer 
\lcdm~to a null measurement.

This significance test 
assumes Gaussian uncertainties, which 
is a good approximation to the distribution of our bootstrap samples.
Since the expected measurement implicitly assumes some model parameters 
(namely, the \lcdm~cosmological parameters) this likelihood ratio 
is equivalent to a Bayes factor describing the preference of our 
data being described by a \lcdm~$\Omega_{\rm M} = 0.3$ 
cosmology over a null measurement of constant $1.16$ after rescaling. 
We find $K = 4.5$, which while not a very strong rejection of null 
due to the large error bars, does provide ``substantial'' evidence,
as usually interpreted in the
Bayesian literature~\citep{Jeffreys1961}. 
We do not convert this likelihood ratio
into a significance since that would require additional modeling of the
posterior shape.
Our individual $\chi^2$ measures are less than one, since our 
error bar estimates should be seen as conservative.

\begin{figure}
  \centering
  {\includegraphics[type=png,ext=.png,read=.png,width=\columnwidth]{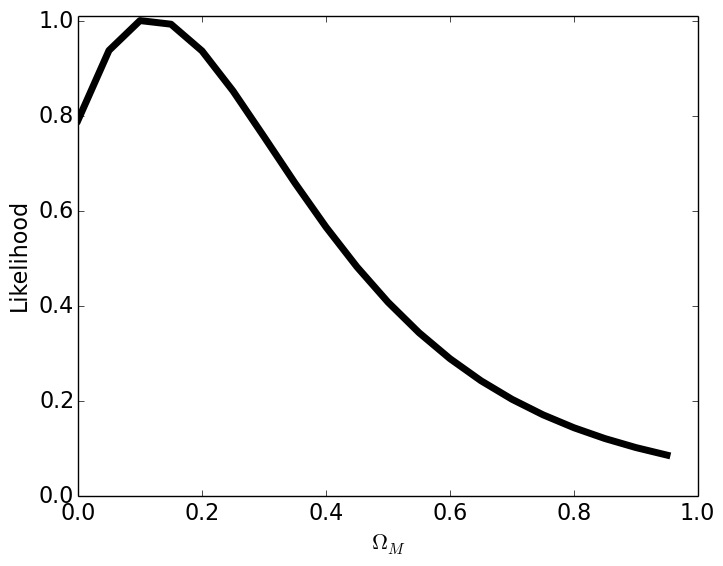}}
  \caption{
           One-dimensional relative likelihood as a function of 
           $\Omega_{\rm M}$ for our stacked void measurements, after
           correcting for peculiar velocities.
          }
\label{fig:drall_likelihood}
\end{figure}

We may also evaluate our measurement by performing a likelihood 
analysis of various cosmologies assuming the above Gaussian likelihood 
function.
Figure~\ref{fig:drall_likelihood} shows the relative likelihood
of $\Omega_{\rm M}$ values in a flat universe with a cosmological
constant, given our stretch measurements.
We calculate this likelihood using the weighted average
measurements in each redshift bin after correcting for the 
effects of peculiar velocities.  
While our error bars do not allow a precise measurement of 
$\Omega_{\rm M}$, we can read off likelihood ratios for various 
cosmologies. Most significantly, we disfavor an $\Omega_{\rm M} = 1.0$ 
universe by a factor of $\sim$10, which corresponds to substantial
evidence for the AP effect expected in the LCDM model.

% -----------------------------------------------------------------------------
% -----------------------------------------------------------------------------
\section{Conclusions}
\label{sec:conclusions}

We have used $\sim$1,500 voids identified with the watershed transform 
code {\tt VIDE}
in the SDSS DR7 main sample and BOSS DR10 
LOWZ and CMASS samples to perform an \ap~test. We stacked voids 
to construct standard spheres and measured the ratio of their 
line-of-sight to angular extents, or \emph{stretch}. 
We used voids found in mock populations with 
realistic number densities and clustering properties to assess the 
impact of peculiar velocities and optimize the stacking to produce 
the most significant result. After correcting for systematic effects 
and measuring the void stretch from redshift $z=0.05$ to $0.6$ 
we translated our \ap~estimation into a constraint on $\Omega_{\rm M}$.
We find a best fit value of $\Omega_{\rm M} ~\sim 0.15$, and our measurements 
prefer this value over $\Omega_{\rm M} = 1.0$ by a factor of 10.
We find a likelihood ratio of 4.5 for our results to reject a null 
measurement.
Taken together,  
we interpret these as a substantial detection of the \ap~test with our 
sample of cosmic voids.

We have verified the uniform and constant systematic offset 
caused by peculiar velocities originally seen in the pure 
$N$-body simulations of~\citet{LavauxGuilhem2011} to also 
apply to realistic galaxy populations. This 14\% line-of-sight 
flattening appears 
universal for all void sizes studied (7-80~\hmpc), 
all redshifts studied ($z=0.0$-$1.2$), and all tracer densities 
studied ($3 \times 10^{-4}$ - $1.0$ particles per cubic~\hmpc). 
We observe this flattening regardless of the composition of the void 
stack, once a minimum threshold number of voids is met. 
Indeed, the AP measurement is quite binary: either no signal 
is obtained at all (if there are too few voids) or the measured 
signal has the expected uniform distortion.
While we have some preliminary indication as to the source 
of this offset, we relegate a full analysis of the effects of peculiar 
velocities on voids found using {\tt VIDE} to a forthcoming paper.

We used our mocks to find the minimum number of voids necessary 
in a stack to obtain a measurement and used these results 
to optimize our result in data. We did not perform an exhaustive 
search through the space of \emph{all} possible 
configurations (e.g., optimal number of redshift 
bins, volume-weighted samples, stacking configurations, etc.) which 
leaves open the possibility of a substantial improved measurement 
with current data.

Our previous application of the AP test to cosmic voids~\citep{Sutter2012b} 
did not correct for systematic effects but did marginalize over potential 
bias values in the final likelihood analysis, which explains the very large 
uncertainty and preference for higher $\Omega_{\rm M}$ in that analysis. 

Based on dark matter simulations~\citet{LavauxGuilhem2011} 
predicted that in the full BOSS survey 
an AP analysis with voids would be competitive with BAO measurements 
from the same survey set. At this stage our Bayes factor of 4.5 
presents substantial but not yet strong or decisive evidence for the 
AP effect in the current BOSS void sample.  
Several factors contribute to this difference. First, we do not yet have access to the full BOSS survey, which will include more galaxies within 
the same survey footprint, increasing the number density and hence 
accessing a much larger number of small voids.
Second,~\citet{LavauxGuilhem2011} provided forecasts using a profile-based 
shape measurement technique.
We found that this method results in far better ellipticity measurements 
but only when the number of voids in a stack exceeds a 
threshold of $\sim$100 for dense surveys and $\sim$300 for sparse 
surveys within relatively narrow radius bins --- otherwise the 
fit fails catastrophically with 
high probability. The redshift width of our volume-limited samples 
prevents us from forming stacks of the required number of voids.
Finally, the earlier study used extrapolated abundances from voids found 
in the dark matter particle distribution rather than the more 
realistic simulations of voids found using dark matter halos 
or HOD galaxies as 
tracers~\citep[e.g,][]{Furlanetto2006, Jennings2013, Sutter2013a}.

Our analysis of the AP 
test in void catalogs drawn from realistic mocks shows that the data quality is about to cross a threshold where the AP test based on stacked voids will  yield competitive and complementary measurements to those based on BAO.
All that is needed is more voids to enhance the 
signal-to-noise, add more independent stacks, and allow
measurements at higher redshifts. The BOSS survey itself will provide 
more voids with upcoming data releases, and future spectroscopic surveys 
such as WFIRST~\citep{Spergel2013}, Euclid~\citep{Laureijs2011}, 
or the Square Kilometer Array~\citep{Jarvis2007} will dramatically 
increase the number of known voids from thousands to millions, 
allowing this analysis to move from detection of the effects 
to precision constraints and measurements of fundamental cosmological parameters.

\section*{Acknowledgments}
The authors acknowledge
support from NSF Grant NSF AST 09-08693 ARRA. BDW
acknowledges funding from an ANR Chaire d'Excellence (ANR-10-CEXC-004-01),
the UPMC Chaire Internationale in Theoretical Cosmology, and NSF grants AST-0908
902 and AST-0708849.
This work made in the ILP LABEX (under reference ANR-10-LABX-63) was supported by French state funds managed by the ANR within the Investissements d'Avenir programme under reference ANR-11-IDEX-0004-02.
The authors thank Qingqing Mao for suggesting the ellipticity 
measurement method and Nico Hamaus, Guilhem Lavaux, and David Weinberg
for useful comments and discussion.

\footnotesize{
  \bibliographystyle{mn2e}
  \bibliography{studyap}
}

\end{document}